\documentclass{aastex6}

\newcommand{\ha}{H$\alpha$}
\newcommand{\hb}{H$\beta$}
\newcommand{\othree}{[O~III]}
\newcommand{\stwo}{[S~II]}

\begin{document}


\title{The Aesthetics of Astrophysics:  How to Make Appealing Color-Composite Images that Convey the Science}


%
%
\author{Travis A. Rector\altaffilmark{1,2}}
\affil{Department of Physics \& Astronomy, University of Alaska Anchorage, 3211 Providence Dr., Anchorage, AK 99508, USA}

\author{Zoltan G. Levay and Lisa M. Frattare}
\affil{Space Telescope Science Institute, 3700 San Martin Drive, Baltimore, MD 21218, USA}


\author{Kimberly K. Arcand and Megan Watzke}
\affil{Smithsonian Astrophysical Observatory, 60 Garden St., Cambridge, MA 02138, USA}



\altaffiltext{1}{Gemini Observatory, 670 N. A'ohoku Place, Hilo, HI 96720}
\altaffiltext{2}{Visiting astronomer, Kitt Peak National Observatory, National Optical Astronomy Observatory, which is operated by the Association of Universities for Research in Astronomy (AURA) under a cooperative agreement with the National Science Foundation.}

\begin{abstract}

Astronomy has a rich tradition of using color photography and imaging, for visualization in research as well as for sharing scientific discoveries in formal and informal education settings (i.e., for ``public outreach.")  In the modern era, astronomical research has benefitted tremendously from electronic cameras that allow data and images to be generated and analyzed in a purely digital form with a level of precision not previously possible.   Advances in image-processing software have also enabled color-composite images to be made in ways much more complex than with darkroom techniques, not only at optical wavelengths but across the electromagnetic spectrum.  And the internet has made it possible to rapidly disseminate these images to eager audiences.

Alongside these technological advances, there have been gains in understanding how to make images that are scientifically illustrative as well as aesthetically pleasing.  Studies have also given insights on how the public interprets astronomical images, and how that can be different than professional astronomers.  An understanding of these differences will help in the creation of images that are meaningful to both groups.

In this invited review we discuss the techniques behind making color-composite images as well as examine the factors one should consider when doing so, whether for data visualization or public consumption.  We also provide a brief history of astronomical imaging with a focus on the origins of the ``modern era" during which distribution of high-quality astronomical images to the public is a part of nearly every professional observatory's public outreach.  We review relevant research into the expectations and misconceptions that often affect the public's interpretation of these images.


\end{abstract}




\section{Introduction} \label{sec:intro}

Imagery plays an important, yet complex, role in helping to communicate science \citep{trumbo2000}. Images are relevant across science disciplines \citep{frankel2004,nicholson2005}, and particularly in astronomy \citep{kessler2012}. From photographs of issues involving climate change, to medical scans of patients, to visualizations in scanning electron microscopy, to snapshots of the Universe, such images don't qualify as ``records of the real," \citep{rothstein2010} but rather are representative  \citep{smith2016}.  This is especially true in astronomy, where images are used for research as well as a means of sharing scientific discoveries in formal and informal education settings.

\subsection{Early History}

Photography has played a vital role in astronomy for more than 150 years. In 1840, American chemist John William Draper took the first successful photograph of the Moon. In 1880 his son, Henry Draper, was the first to take an image of a deep sky object, M42. By the end of the 19th century, the development of photographic plates had advanced to the point where they were the primary tool for astronomical research, a role they served for roughly 100 years.  

Photographic plates, like modern-day charge-coupled devices (CCDs), produce only ``black and white" (i.e., monochrome or grayscale) images.  The three-color method is the process by which nearly all color images, astronomical or otherwise, are made from these grayscale images.  First described by \citet{maxwell1885},  the method works on the principle that it is possible to create every color visible to the human eye simply by combining different relative amounts of red, green, and blue light.  Thus, by combining grayscale images taken through red, green, and blue filters, either through chemical or electronic means, the ``photographer" can create a full-color image by using only these three bands of light.   

During the second half of the 20th century, scientists and photographers such as Bill Miller at Palomar Observatory, David Malin at the Anglo Australian Observatory (AAO), and Mark Hanna at the National Optical Astronomy Observatory (NOAO), mastered the art of creating color astronomical images by using the three-color process with photographic plates.  These imaging pioneers also developed innovative darkroom techniques to maximize the amount of detail visible in the images, a precursor to the sorts of tools now available in digital image-processing (IP) software such as Adobe Photoshop\texttrademark. 

By the 1980s technology had advanced to the point where electronic (e.g., CCD) cameras were being used in earnest on many professional telescopes for imaging purposes.  Despite the advantages of CCDs, many astronomers continued to use photographic plates because of their larger field of view (FOV). Whereas photographic plates were roughly the size of a sheet of paper, the first CCDs were smaller than a postage stamp. Modern CCD detectors are still smaller than a photographic plate, but newer electronic cameras employ arrays of multiple CCDs to cover a larger FOV. The use of photographic plates waned in the 1990s. And by 2000 the production of photographic plates was discontinued.

\subsection{Three Key Developments}

The development of CCDs was essential for the digital manipulation of astronomical images.  It was also essential for putting telescopes into space. The Hubble Space Telescope (HST), jointly run by NASA and the European Space Agency (ESA), is  now synonymous with beautiful images.  But it is easy to forget that Hubble's performance was initially flawed due to miscalculations in the shape of the mirror prior to launch. Fortunately the spherical aberration in the primary mirror was reparable with corrective optics, which were installed in December 1993.  Images from WFPC2, while not quite to the level of quality to which we are accustomed from Hubble today,  nonetheless showed stars and galaxies with a sharpness not seen before by Earth-bound telescopes. The images effectively counteracted the bad publicity that HST and NASA initially endured. In a peculiar twist, the spherical aberration helped lay the groundwork for creating and sharing astronomical images in a way not done before but that we take for granted today.  For example, it led to the establishment of the STScI News Office, which prepares and distributes high-quality HST images, graphics, and text as news releases.  Many professional observatories also now have a news office or an ``education and public outreach" (EPO) office that do the same.  It also led to the development of deconvolution software to correct for the aberration computationally, which was effective in improving the qualitative appearance of the images in addition to providing better analytical results.

\begin{figure}
\figurenum{1}
\plottwo{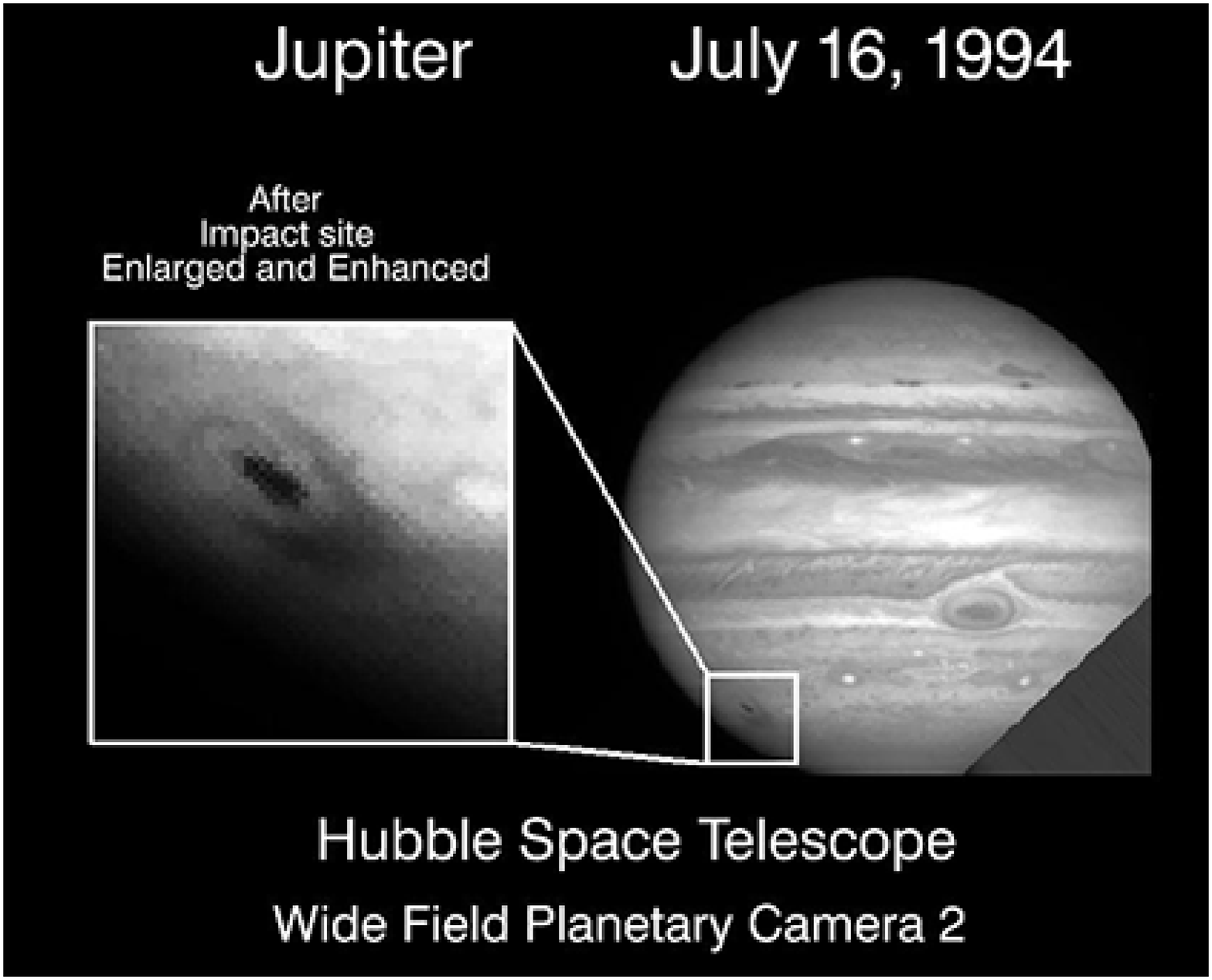}{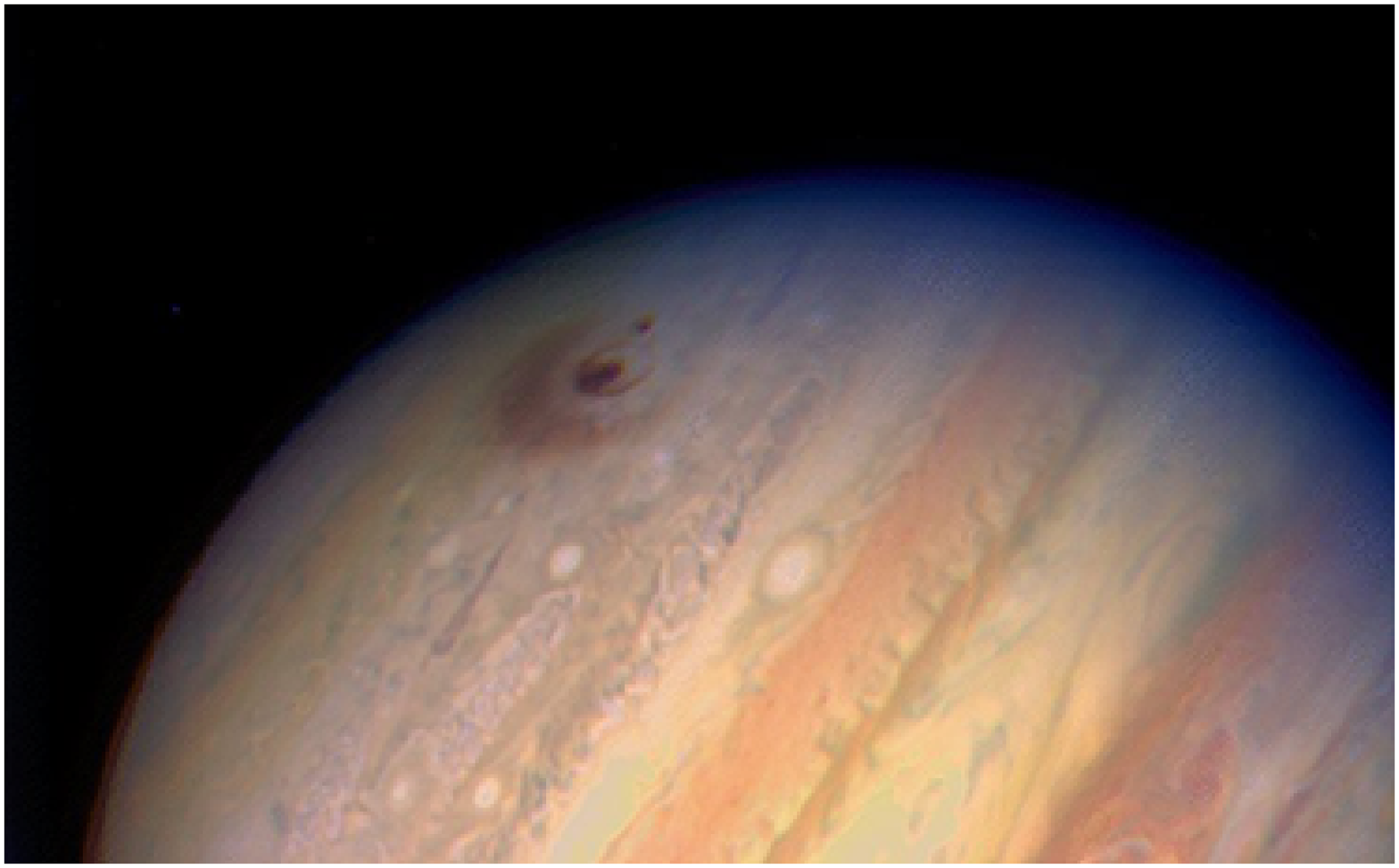} 
\caption{These HST images \citep{nasa1994a,nasa1994b} show the impact from the first fragment of comet Shoemaker-Levy 9 with Jupiter.  They are primitive in comparison to modern HST image releases but nonetheless were popular because, for the first time, the World Wide Web facilitated the rapid distribution of images from an exciting astronomical event. The image on the left was released a day after the impact. \label{fig:f1}}
\end{figure}

The repair of HST was timely, as it would be ready to observe the comet Shoemaker-Levy~9 (SL9) impact with Jupiter in July~1994. Meanwhile, another technological development had recently occurred that would also prove to be important for this event. The World Wide Web, the standard for Internet communication today, matured in 1993 with the introduction of the Mosaic  web browser from the National Center for Supercomputing Applications. While still in its infancy in 1994, the Web would prove to be an effective way to rapidly distribute images of the SL9 impact to the public, e.g., those in Figure~\ref{fig:f1}. Despite problems with long download times and crashed servers, the SL9 impact demonstrated the power of the Web for quickly disseminating astronomical images and information. It also illustrated that there was a tremendous demand for such astronomical images.


A third development was also important. At the same time that electronic camera hardware was improving, so was the software. IP programs, such as Photoshop, quickly became essential tools for creating and modifying electronic images. Sharing a coupled history, the first version of Photoshop launched in the same year as HST.  In September 1994, a new version of Photoshop (v3.0) added a feature that would prove invaluable for the creation of more complex color images. While not the first program to do so, Photoshop~3.0 had the ability to use ``layers."

In the layering metaphor, multiple images can be combined to form a single image. In the traditional three-color method, three grayscale images were colorized red, green, and blue, and then combined to form the color image. While simple and powerful, this method is limited in that it uses only three images and only these three colors. With the new layering capabilities available in Photoshop$^1$\footnotetext{And in other programs, such as the GNU Image Manipulation Program (GIMP): https://www.gimp.org}, it was now possible to combine any number of images and to use colors other than red, green, and blue. The ability to put many (3+) layers of astronomical data together with different colors was a critical technological development for creating many of the spectacular images we enjoy today.  Such images are now quite common.  For example, a color-composite image of the Hubble Ultra Deep Field uses thirteen datasets \citep{nasa2014}.  

\subsection{The Modern Era}

A year after the SL9 impact, another event would play an important role in the development of astronomical imaging. In November 1995, NASA released the iconic image of the center of M16, now informally known as the ``Pillars of Creation" \citep{hester1996}. The striking nature of the image, combined with the evolving Internet's ability to deliver images directly to individuals, helped to make it hugely popular with the general public. For example, the New York Times described it as a ``remarkable view of a kind of fantasyland of powerful cosmic forces" \citep{wilford1995}.   At that time receiving national news attention for an astronomical image was rare and notable.

The popularity of the Pillars of Creation and other HST images served as the inspiration to start a program where HST data, collected primarily for scientific purposes, would be used to create beautiful astronomical pictures for public outreach. Since 1998, the Hubble Heritage program has released a new color image nearly every month. The project has produced some of HST's most popular images. 

Following in Hubble's footsteps, other observatories began to release high-quality color images for public consumption. Also in 1998, NOAO started to use its large-format Mosaic cameras to produce color images for the public. Starting in 2002, the Canada-France-Hawai'i Telescope has their ``Hawaiian Starlight" program \citep{cuillandre2002}. And in 2004, Gemini Observatory started its Gemini Legacy image project.  Most space telescopes launched since HST have integrated color image creation and dissemination into their programs as well. NASA's three remaining Great Observatories, HST, Chandra, and Spitzer, also frequently work together to make images of astronomical objects over a wide portion of the electromagnetic spectrum.  Nowadays nearly every major observatory has a program or process in place to release color astronomical images for outreach purposes.  While somewhat controversial at first, most scientists now see these programs as an effective way to increase interest in astronomy and in science in general.

\section{Expectations and Misconceptions in Interpretation} \label{sec:exp}

When creating color astronomical images it is important to be mindful of how people interact with them.  When first viewing a new astronomical image, a primary concern for many is its veracity.  Commonly asked questions by the public include ``Is this image real?", ``Is this what it really looks like?", or even ``If I were standing right next to this, is this what I would see?"  \citep{rector2015}.  The first question pertains to the existence of the object, which is relatively simple to answer.  The next two questions instead pertain to how we might see it, especially from a closer vantage point.  These questions often reflect misunderstandings of how telescopes, and our eyes, work.  

It is helpful to first consider how our eyes (and brain) see.  There are two types of cells in our eyes that detect light: rods and cones. There are about 90 million rods cells in the human eye, compared to only about 4.5 million cones \citep{curcio1990}. Rods are sensitive to faint light and are concentrated along the outer edges of the retina, which is why our peripheral vision is better at night. The rods however don't perceive color. Instead we see color with the cones. We have three kinds of cones in our eyes: S-cones, M-cones, and L-cones. The names reflect that they are sensitive to short, medium, and long-wavelength visible light respectively. S-cones are primarily sensitive to blue light. The M and L cones are both sensitive to green light. And the L cones are most sensitive to red light. Cones aren't as sensitive as the rods, so they only work in brighter light. Only a handful of stars, e.g., Betelgeuse, are bright enough to activate the cones so that we can see their color.  The brain interprets and combines the signals from these three types of cones to determine perceived color. Indeed, this is why the three-color process is able to reproduce the full gamut of colors.$^{2}$\footnotetext{There are physiological variations in some people. Color blindness is the condition of missing one of the cone types; e.g., protanopia is when a person's eyes lack L cones and cannot see red. Color blindness affects less than one percent of the population; and it primarily affects men.}

Many people perceive the purpose of a telescope as merely to magnify our human vision; i.e., to make a small object look bigger (or to make a more distant object look closer.)  They are unaware that a telescope also amplifies the faint light from an object, as well as allows us to detect light beyond the optical window.  People often assume that astronomical objects are ``too small to see" without a telescope and are surprised to learn that many have angular sizes comparable to or larger than the Moon.  They are also surprised to learn that the surface brightness of an extended object, such as a galaxy or nebula, does not depend on distance in the local Universe; i.e., if you can't already see it from Earth you won't be able to see it even if you are much closer. 

The fact that most non-stellar astronomical objects, with the exception of those within our Solar System, are essentially invisible to the human eye complicates what it means to say ``what it truly looks like."  An essential point to make when discussing astronomical images with the public is that telescopes give us, in effect, superhuman vision.  They literally make the invisible visible.  This is sensible upon reflection, as what would the purpose be of building a large, expensive telescope if it showed us the same view as our own eyes?  Astronomical images are therefore a translation of what the telescope can see into something our eyes can see- and, just as important, our minds can understand.  Because objects in space are fundamentally exotic and unfamiliar, we can use knowledge about the psychology of vision to help us create images that can be naturally interpreted by the viewer.

For example, how people see and interpret color depends on several factors, including cultural variations and expertise.  In general, people will agree upon the basic colors. What is red to one person will be red to many people. There are some cultural variations in the names of colors, and to what range of colors the name applies. For example, there are eleven primary color words in the English language. Other languages have more or fewer; e.g., see \citet{kay1999} for a discussion of the development and use of color in different languages.

When we look at an image, our brain will attempt to derive information from the colors.  This too will depend somewhat on the culture from which we come. For example, red is considered a sign of luck, happiness, and prosperity in China, whereas in Western cultures it is often associated with anger and danger. We also use color to discern physical properties of an object. For example, we often associate color with temperature. Because of  fire, we naturally think of red as hot.  And most people  tend to associate the color blue with cold because liquid water and ice tend to be bluish. It is indeed ironic then that bluer stars are hotter. For this reason, how you associate color with temperature depends somewhat on your knowledge of physics and astronomy. Experts are much more likely to picture blue as hot than are novices; about 80 percent of novices see red as hot compared to 60 percent of experts \citep{smith2015}.

Another way we use color is to perceive depth. We subconsciously interpret color to estimate the relative distance to objects far away. Here on Earth, scattered sunlight from air molecules give distant objects a bluish tint. More distant objects have more intervening air and therefore look bluer. Without this effect, distances are harder for us to visualize. For example, Apollo astronauts when on the moon commented that they had difficulty visually estimating the distance to mountains and hills.  Because of this effect we  naturally interpret blues and greens (collectively known as cool colors) as being more distant to yellows and reds (known as warm colors). We naturally infer depth in images based upon this optical illusion. We do this when we look at astronomical pictures too.  This has led many to ask how telescopes are able to see in 3D.  In the case of HST, this is connected to a common misconception that it actually ``flies" through space to be closer to the targets it observes, a fallacy reinforced by the fact that its solar panels look like wings. The 3D misconception is instead a result of the innovative use of color in many Hubble images, not the inherent advantages of the telescope itself.$^3$\footnotetext{In fact, many non-Hubble astronomical images are assumed to be from HST because they use color in a similar way.}

Our brains also use other visual cues to perceive depth within an image.  Of particular relevance to astronomical images is an effect known as ``size constancy" \citep{boring1964}.  If an image has two objects of similar appearance (e.g., evergreen trees), the brain will interpret the smaller one to be further away. Similarly, the distance to stars of similar color will be inferred by their apparent size in the image, although this is of course instead a consequence of their apparent magnitude.  This effect can give some astronomical images an appealing sensation of depth to them.


While the focus of this paper is on images, we note the importance of high-quality accompanying text.  While it is true that images can ``overpower words" \citep{lazard2014}, astronomy images often require context to be understood.  Images accompanied by a well-written, jargon-free narrative that tells a story can help make the science more attractive \citep{smith2011,smith2015}. Also effective is the use of analogies and factors of relevancy to everyday life to help explain complex or foreign subject matter, and leading questions that draw the viewer into wanting to know more. \citep{smith2016}.  Unfortunately many astronomical images are distributed via social media with no explanation or text (not to mention credits or links to the original source), which can lead to confusion or misconceptions about the content of the image.  

There is also an increase in the number of ``fraudulent" astronomical images distributed on social media.  These are images created by improperly modifying or combining real images (e.g., superimposing a full moon onto a western horizon at sunset), by using computer-generated imagery (CGI) to create the appearance of natural phenomena (e.g., simulating the aurora), or both.  These fake images distract attention from, and cause an increased suspicion of, authentic astronomical images.  In a world where many images are ``photoshopped" (a word that colloquially now means an image has been unethically modified), the more fantastic the image the less likely it is to be trusted.

\section{Technique} \label{sec:tech}

To produce a color image it is necessary to have at least two datasets, and preferably three or more. A dataset is defined as a two-dimensional image of a particular waveband, polarization, or other distinct characteristic; e.g., an optical image through a single filter, a radio image at or in a particular waveband and/or polarization, or an X-ray image over a specified energy range.  The data must first be fully reduced, e.g., optical data are bias and flat-field corrected, with a standard data-reduction package. Data from multiple instruments and/or telescopes should be projected to a common world coordinate system (WCS).

\citet{rector2007} offer a detailed explanation for the steps necessary to create a high-quality color-composite image from this data.  In brief, the process can be distilled into the following steps: (1) convert (or ``project") each dataset into a grayscale image; (2) import these images as layers into an IP software package; (3) adjust the intensity scaling of each layer to better show detail (i.e., increase contrast in regions of interest); (4) assign a color to each layer; (5) fine tune the image, which includes overall color balance, the removal of residual artifacts, and orientation and framing; and (6) prepare the image for electronic distribution and print production.  While presented as a linear step-by-step process, it is important to emphasize it is iterative in nature \citep{mazieres2005}.  At many points in the process it may be necessary or desirable to return to previous steps; e.g., to adjust the scaling of an individual layer to improve the overall color balance of the image.

\subsection{Recent Advances in Image Processing}

Since \citet{rector2007} the continued development of IP software has provided new features that allow for more sophisticated techniques as well as a more streamlined workflow.  Some of these new features enable the creation of images in innovative ways.  Others relate to applying adjustments to an image more effectively.  Yet other features relate more to the software user interface, managing aspects of the development workflow more efficiently and flexibly, particularly when working with large and/or complex documents. For example, traditionally the ``screen" algorithm has been used to combine the layers in an image.  However, newer blending algorithms can also be useful, such as the ``multiply" algorithm (which helps to show detail in bright areas) and the ``blend if" option (which helps to combine a selected range of tones).  

Interchangeable color modes in IP software provide opportunities to develop images in new ways.  In particular, the ability to decompose luminosity from color allows different datasets to be used for each component.  For example, it is now possible to use layer blend modes in Photoshop to combine a high-resolution (e.g., HST) monochrome image as a ``luminosity" (i.e., grayscale) image with a lower-resolution color-composite image generated from ground-based telescopic data.  An illustration of this technique is a wide-field mosaic of the Carina Nebula (NGC 3372) that combines a single-band mosaic of HST images generated from ACS/WFC \ha\ data, rendered in luminosity, with a composite of CTIO 4m MOSAIC narrow-band images (\ha, \stwo, and \othree) rendered in color \citep{levay2007}.  The largest difference between the images is spatial resolution, $\sim$0.1 arcsec for HST as compared to $\sim$1 arcsec for CTIO.  This technique effectively generates an image with the sharpness of ACS without requiring all of the data be obtained with HST.

Datasets disparate in resolution can also be used to improve a composition by extending a narrow field of view or to fill in gaps or irregular areas.  One or more images with a small field of view at high resolution can be combined with a wide-field  image at lower resolution to produce a more pleasing composition, provided the datasets are compatible.  Adjustments may be necessary to match the colors and tonalities enough to combine the images.  An example is an image of the Helix Nebula (NGC 7293) generated from HST ACS and CTIO 4m MOSAIC data \citep{odell2004}.  

The large dynamic range in brightness of some astronomical objects may require multiple projections of a dataset to effectively show detail in bright {\it and} dark regions in a single composition.  In some cases the brighter areas of a target may be entirely saturated in order to obtain adequate signal in fainter regions.  In this case additional shorter exposures may be necessary to avoid saturating the bright regions.  In either case the two renditions of the image can be combined using ``layer masks" to select the optimal areas of each image.  The HST/CTIO mosaic of the Carina Nebula is also an example of this technique.

When setting up observations of an object, a judicious selection of filters or bandpasses can result in a more natural or pleasing result.  For example, narrow-band data of nebular targets can effectively render the various emission-line components in vivid, well-separated colors.  However continuum sources such as stars in the image often lose their expected colors because the narrow-band filters are not chromatically ordered or do not adequately sample the blackbody stellar spectra.  Adding broad-band filters to the image composition will reproduce the colors of stars more naturally.  In general the exposure times for the broadband filters can be shorter, thus their contribution in signal to the nebulosity will be relatively small.   Separate color images may be combined from the narrow-band and broad-band filter sets and then superimposed, or a single image can be generated from all of the datasets.  

As images become larger and more complex, not only in number of pixels but also in number of datasets used, effective management of the workflow will be even more important.  The use of adjustment layers, smart objects, and linked image layers allows filters and adjustments to be applied to an image non-destructively.
It also can streamline the workflow where multiple layers in the document behave as a single pixel layer, but remain editable.  A powerful use of this feature is to enable editing of a small version of a document including resizing, rotating, applying adjustments and filters, while preserving all the pixel resolution in its constituent images.  

Various methods may be used in managing multiple layers in editing software.  In Photoshop, layers may be combined into groups, allowing a set of layers to be treated as a single logical layer, with the option of changing the layer blend mode, adding masks, and and other attributes otherwise applied to layers. This can be an effective alternate to using clipping masks to apply adjustments to an image layer, e.g., in applying color to a monochromatic layer. 

\citet{rector2007} discusses common telescope and CCD artifacts, such as CCD charge bleeds, diffraction spikes, optical ghosts, and cosmic rays, along with a recommended treatment. Many of the defects on this list are dealt with after the observation has been made, with cosmetic removal or adjustment tools in the IP software. Naturally it is worthwhile to anticipate such artifacts and schedule observations to avoid distracting artifacts {\it a priori} if possible. Examples include rotating the FOV for optical CCDs so that charge bleeds do not intersect regions of interest.  Or adding ``postflash" exposures to eliminate ghost images from bright stars in previous exposures.  Or using many half-field exposures to help with blending multiple pointings taken with instruments that have chip gaps.  It is important to emphasize that IP tools that modify the actual content of an image should only be used to remove cosmetic defects within the image.  Whether an image is intended for scientific visualization or public consumption, there is an expectation that the image is an accurate representation of the data.  Inappropriate modification of an image represents a violation of that trust, particularly when submitted to a refereed journal \citep{rossner2004}.  For this reason, many journals have guidelines as to what image modifications are acceptable, e.g., \citet{nature2006}.

\subsection{Color Schemes}

\citet{rector2007} explain how different color schemes can be used when creating an image, with examples.  Several terms have come into usage to describe these color schemes.  Unfortunately these phrases are not always used in a consistent manner.  The phrase ``true color" is commonly used when an image is created using broadband optical filters that roughly match the sensitivity of the cones inside our eyes. But these images do not necessarily correspond to  what our eyes would see.  Because the human eye's sensitivity to color varies with intensity of light, ``true color" is complex when discussing faint objects.$^4$\footnotetext{For bright, sunlit Solar System objects, true color images that show ``what your eyes would see" are possible using color calibration targets, such as the sundials mounted on the NASA Mars Spirit and Opportunity rovers.} For example, if you look at M42 through a telescope it will look cyan because our eyes more easily detect the \hb\ and \othree\ emission than the more dominant \ha\ emission, to which our eyes are not sensitive.  Perhaps the only valid true-color image for most galaxies and nebulae is to show a field of black, with only a handful of stars visible \citep{rector2015}.  For this reason it is recommended to avoid using this phrase.

Equally confusing are the terms Òfalse colorÓ and Òpseudo color.Ó They sound like they're synonymous, but they refer to different ways of using color. These terms aren't always used consistently, but they do have distinct meaning. The term false color is often used when referring to an image that was made with two or more filters, but the colors assigned to the filters are not close to the ``intrinsic" color of the filter; i.e., the color it would appear to be if viewed against a bright, white light (assuming it is an optical filter).  An example of false color would be if you assigned blue to a filter that looks orange to your eyes. While rarely, if ever, done with broadband filters, it is common practice with narrowband filters. Unfortunately the phrase implies that the objects in the image are themselves not real.  For this reason, the phrase ``representative color" is often used instead because the colors represent the science being shown in the image.

\begin{figure}
\figurenum{2}
\plottwo{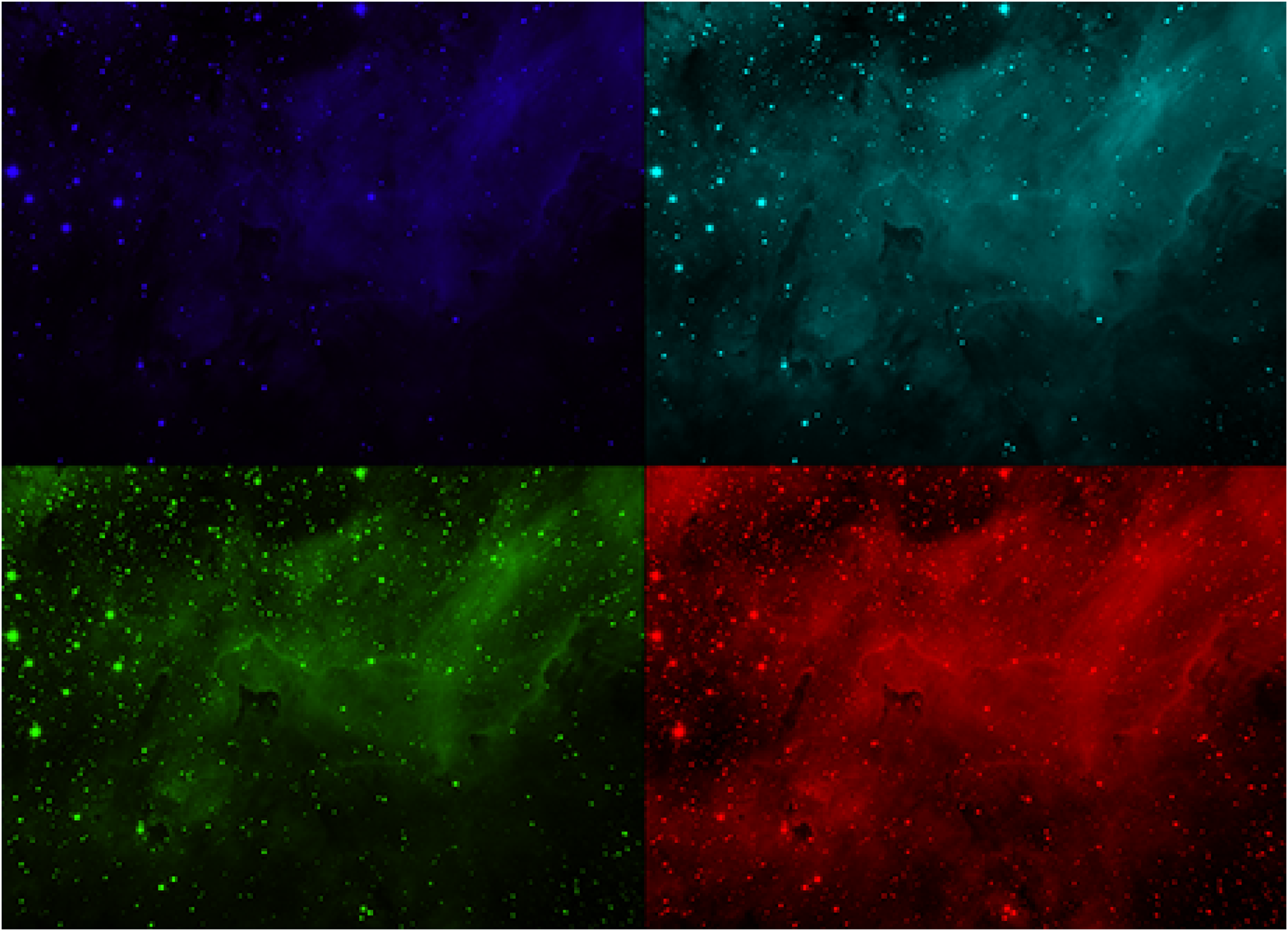}{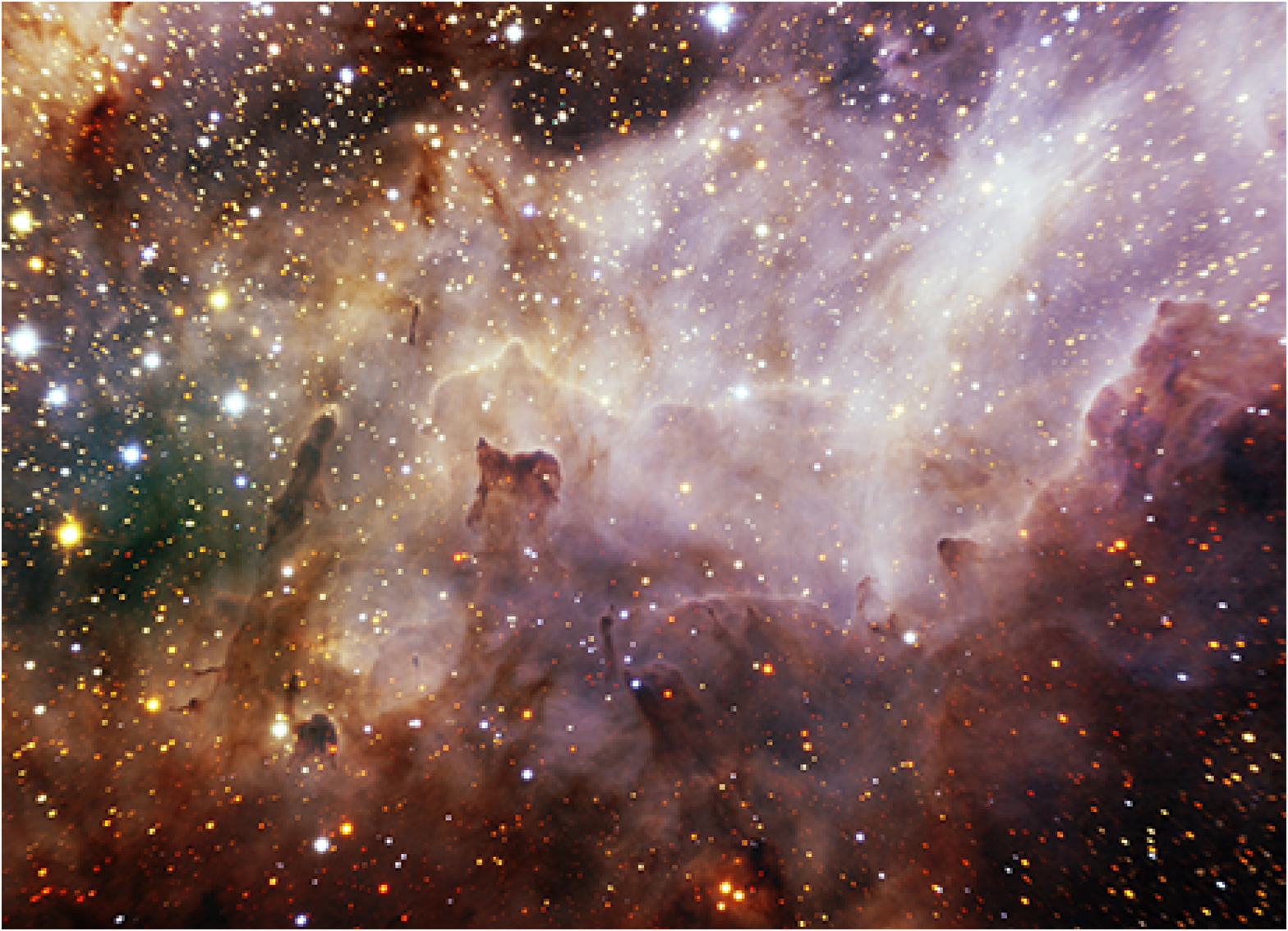}
\caption{Released by \citet{gemini2013}, this FLAMINGOS-2 image shows a region of M17 in the near infrared.  The image (right) was created by combining images obtained through $Y$, $J$, $H$, and $K_s$ filters that were colored blue, cyan, green, and red respectively (left).  The filters are therefore arranged in a chromatic ordering.\label{fig:f2}}
\end{figure}

Since in representative color the assigned colors don't match the intrinsic color of the filter, it is essential to think carefully about the colors used. Assigned colors should be chosen primarily to best illustrate what's interesting scientifically. ``Chromatic ordering" is where the filter that corresponds to the lowest-energy light is made red, the filter that corresponds to the highest energy is blue (or violet), and the filter of intermediate energy is made green. If more than three filters are used, other colors might be chosen; e.g., the filter between the red and green might be assigned yellow.  Chromatic ordering is almost always used for broadband filters and energy ranges, regardless of the wavelength regime (see Figure~\ref{fig:f2} for an example.)  Non-chromatic (sometimes called ``composite") color schemes are often used in images made with narrowband data. 

An unrelated color scheme is that of pseudocolor. In a pseudocolor image, color is encoded in a monochromatic image to reflect a physical property; e.g., intensity, polarization, or radial velocity. Pseudocolor images can convey important information in a visually compelling manner, but they can also be confusing because color lacks its normal meaning (i.e., a measure of relative amounts of light over different energy/wavelength ranges.)

Color schemes can be confusing when discussing images created with non-optical data.  For example, the phrasing ``true-color infrared image" is occasionally used despite having an ambiguous meaning.  Representative color is, in a sense, always used when making images with non-visible light because they are, by their very nature, invisible to us.  While false color and true color should be avoided for non-optical images, the terms chromatic and composite ordering should be used when appropriate.  For example, \citet{hurt2010} discusses the confusion that can occur when discussing color in infrared images and offers strategies for counteracting common misconceptions.  


\begin{figure}
\figurenum{3}
\plotone{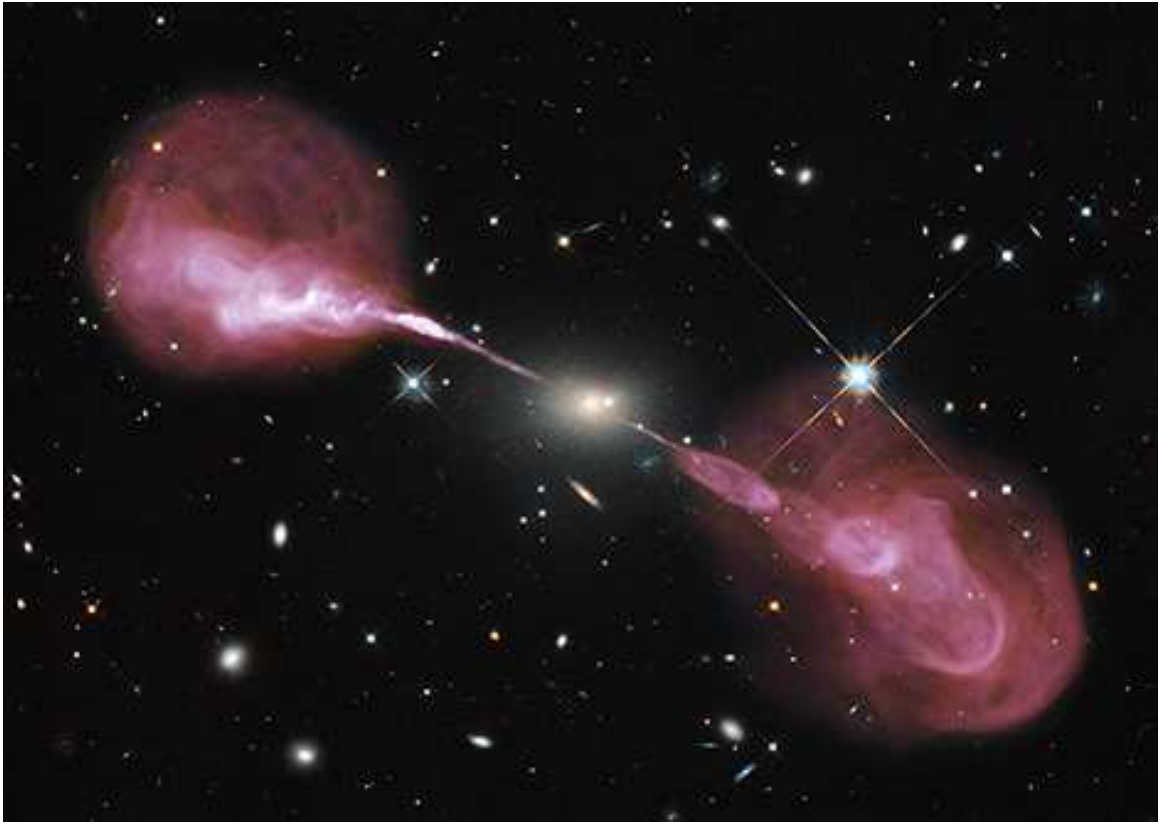}
\caption{This image of the radio galaxy Hercules~A \citep{nasa2012} combines HST WFC3/UVIS data from two optical filters with three radio wavebands obtained with the NRAO Very Large Array telescope.  By using multiple filters/wavebands in each energy range, color information in the radio emission and the natural colors of stars and galaxies in the optical are retained.\label{fig:f3}}
\end{figure}

For multi-wavelength images (sometimes called ``hybrid images"), where data from widely disparate energy bands are combined, it is often better to make a full-color image from each energy band and then combine these full color images rather than to treat each energy band as a monochromatic image \citep{rector2007}.  For example, images that combine radio, optical, and x-ray data often assign red, green, and blue to each energy band.  While this is chromatic ordering, an often desirable color scheme, it causes stars and galaxies (which are often only visible in the optical) to have an unnatural green color.  If possible, it is better to create multiple chromatically ordered color images using filters, bandpasses, or energy slices within each energy range, and then combine these color images to produce the final multi-wavelength composite image; e.g. Figure~\ref{fig:f3}.  This will give stars and galaxies more natural colors as well as reveal color information within the other wavelength bands.  One also needs to be careful when combining imaging data with models; e.g., images of galaxy clusters that combine optical and x-ray data (that show the hot intracluster medium gas) with models of dark-matter distribution (derived from gravitational lensing), e.g., \cite{nasa2006}, can create the misconception that we can actually detect electromagnetic radiation from the dark matter because the dark-matter model is shown in the image as blue ``light."

Unfortunately, there is no set of strict rules that dictate how colors are used in scientific images. It's not possible to do so because there are an infinite number of ways to combine filters and data from different telescopes; e.g., KPNO alone has over 100 filters available for its optical instruments.  The flexibility of current IP software also allows the exploration of many different color schemes in an image simultaneously. It is recommended that the user attempt many different versions before settling on a final image. The ability to combine any number of datasets in a wide range of color combinations opens up an immense parameter space of possibilities.  How then should one go about creating an image that is scientifically valid and useful yet aesthetically pleasing?  In the next section we discuss several factors that can guide the making of an image.

\section{Aesthetic Considerations} \label{sec:aesthetics}

When transforming data into a color image, one must take into consideration not just the data involved, but also the people who may be handling the data and then those ultimately viewing the image.  On one end, there are the experts making choices during the processing of the data, e.g., what datasets to include, how to manipulate the data (e.g., apply smoothing processes), how to crop and rotate the image, and what colors to assign \citep{depasquale2015}. On the other end, there are the people who will be viewing the results.   When creating an image it is essential to consider what you wish for the image to convey.  What meaning people will interpret when viewing the image is just as important question \citep{smith2011}, because the answers to these two questions are not necessarily the same.  The study of the perception and comprehension across the expert to non-expert spectrum of viewers is therefore an important area of study.  

When looking at space images, experts such as astrophysicists start by thinking about the science conveyed in the image and then consider its aesthetic appeal. Non-experts usually go the opposite direction. They focus first on the aesthetic value of an image, but ultimately want to understand the science. Research also shows that non-experts want to know how to understand an image in the way that experts do. This suggests that  the ``full story" should be told to viewing audiences \citep{smith2014a,smith2015}

\citet{lynch1988} argued that images created for scientific purposes are made to maximize detail in the regions of interest, whereas for public outreach a greater emphasis is placed on visual appeal.  However, in both cases subjective decisions must be made in the image-making process.  And these two goals are not mutually exclusive and often overlap \citep{kessler2007}.  Perhaps not surprisingly, many of the principles that make for an appealing piece of art also apply to astronomical image making. Thus it is possible to balance both the scientific and the aesthetic points of view \citep{kessler2012,arcand2013,rector2015}.  Astronomers may be more constrained than artists in what we can do, but we can nonetheless use insights from the art world to help us understand what makes for a good image.

\citet{rector2007} and \citet{christensen2014} discuss several of the factors that affect the aesthetic quality of an image, at least for objects with a large angular size.  Naturally, the quality of the data is important.  People often assume that HST's success lies simply in its high angular resolution.  More important than resolution is the apparent ``sharpness" of the image, which is simply the ratio of the FOV to the angular resolution- a ratio that \citet{christensen2014} call ``photogenic resolution."  For this reason, ground-based optical telescopes with wide-field cameras can actually compete favorably with HST in terms of sharpness.  Whereas high-angular-resolution Very Long Baseline Interferometry (VLBI) observations fare poorly due to the limited FOV.  Other properties of the data, such as signal-to-noise (S/N) and quality of calibration (e.g., flat-fielding) are important.  Calibration errors and other defects, often collectively referred to as ``cosmetic defects," can distract the viewer's attention from the astronomical objects in the image. The proper removal of cosmetic defects from an image is so important that this step in the image-making process often takes the most time. 

When creating images it also is important to consider the size at which the image will be displayed.  For example, it may be (at least initially) seen as a small inset in a newspaper or on a web page (e.g., in a Google search).  Or it may be viewed on the small screen of a mobile device, an increasingly common occurrence.  It is estimated that in the year 2016 about 31 million people in the United States will access the World Wide Web exclusively with their cell phone or other mobile device \citep{emarketer2016}.  While this is only 11\% of all internet users, this fraction is expected to grow.  While more research is needed, it is clear that the device used affects the experience.  Not surprisingly, \citet{smith2014b} found that in general bigger displays are better for making an impact with science visuals. However, this only appears to hold true for larger screens and displays. And \citet{smith2016} found that mobile phone users are more likely to want to learn more about an image as compared to users on desktops, tablets, or laptops. 

\subsection{Color and Composition}

How color and composition are used in the image is just as important, if not more so, than the quality of the data itself.  Section~\ref{sec:exp} delineates several of the ways in which color is naturally used to interpret physical characteristics.  When working with color in an image it is important to consider that our eyes, and indeed our senses in general, function by detecting relative differences; e.g., a circle appears to be larger when smaller circles are present \citep{roberts2005}. A color in an image is similarly intensified or weakened by contrast with other colors present \citep{albers63,itten1970}. Thus, contrasts between colors in an image can be used to highlight or de-emphasize other elements of the image. Different color schemes can also be used to convey different meanings or emotions \citep{whelan94}.  Black and white images have only one contrast scheme, that of light to dark, i.e., the intensity contrast. However, there are seven contrasts in color images \citep{itten1970,itten1990}. Thus, when possible, it is important to use datasets that will maximize the amount of color contrast within the image; e.g., narrowband filters often create strong color contrasts of hue because line emission is sensitive to physical parameters (such as temperature and density) that can vary significantly within the object. \citet{rector2007} offers strategies on how to maximize the color contrasts within an image.  

\begin{figure}
\figurenum{4}
\plotone{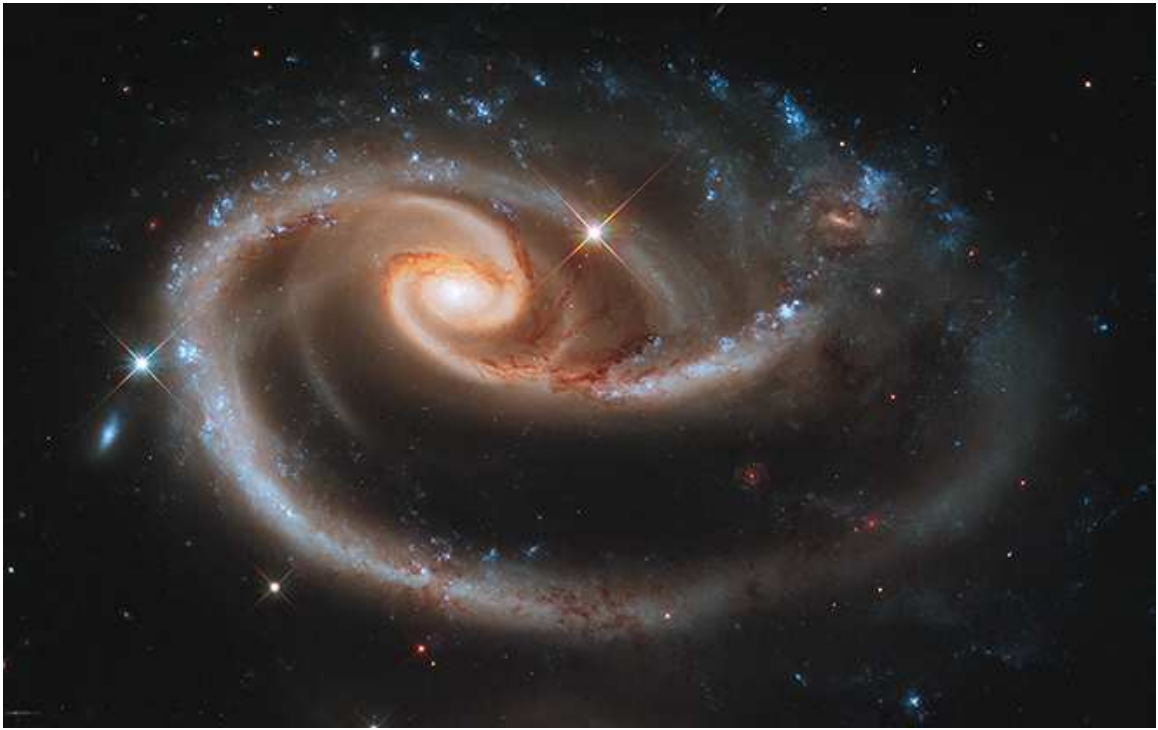}
\caption{A cropped version of an image released by \citet{nasa2011a} shows an HST WFC3/UVIS image of the galaxy UGC~1810.  The symmetry of the inner spiral arms stands in contrast to the distorted, chaotic shape of the outer spiral arms.  The distorted spiral arms naturally convey chaos and motion within the galaxy. \label{fig:f4}}
\end{figure}

In painting, drawing, and photography, composition is the practice of arranging components within the picture to create a certain ``feel" or to draw attention to particular elements.  While we cannot control what is in a particular region of the sky, we can crop and rotate images to affect the composition.  Images are now often cropped to draw attention to the object or areas that are most interesting, scientifically or aesthetically, by using rules of composition; e.g. the rule of thirds and the diagonal method \citep{arnheim1954}.  The orientation of an image can also affect how it is interpreted based upon perceived lines within the image. These lines may be real, such as striations within a nebula. Or they may be imagined, such as an imaginary line between two galaxies.   The orientation of these lines affect the perception of the image, e.g., horizontal lines can create a sense of stability, tranquility and spaciousness; vertical lines can give a perception of size and majesty; diagonal lines can cause a feeling of tension and motion; converging lines can create the illusion of depth; e.g., train tracks converging in the distance; and curved lines can give the sensation of softness as well as gentle movement \citep{grill1990}. Like with color, contrasts in structure can emphasize elements of the image \citep{wong1972}; e.g., Figure~\ref{fig:f4} shows that a departure from the symmetry in the spiral structure of a galaxy will draw attention to the anomaly.

\begin{figure}
\figurenum{5}
\plottwo{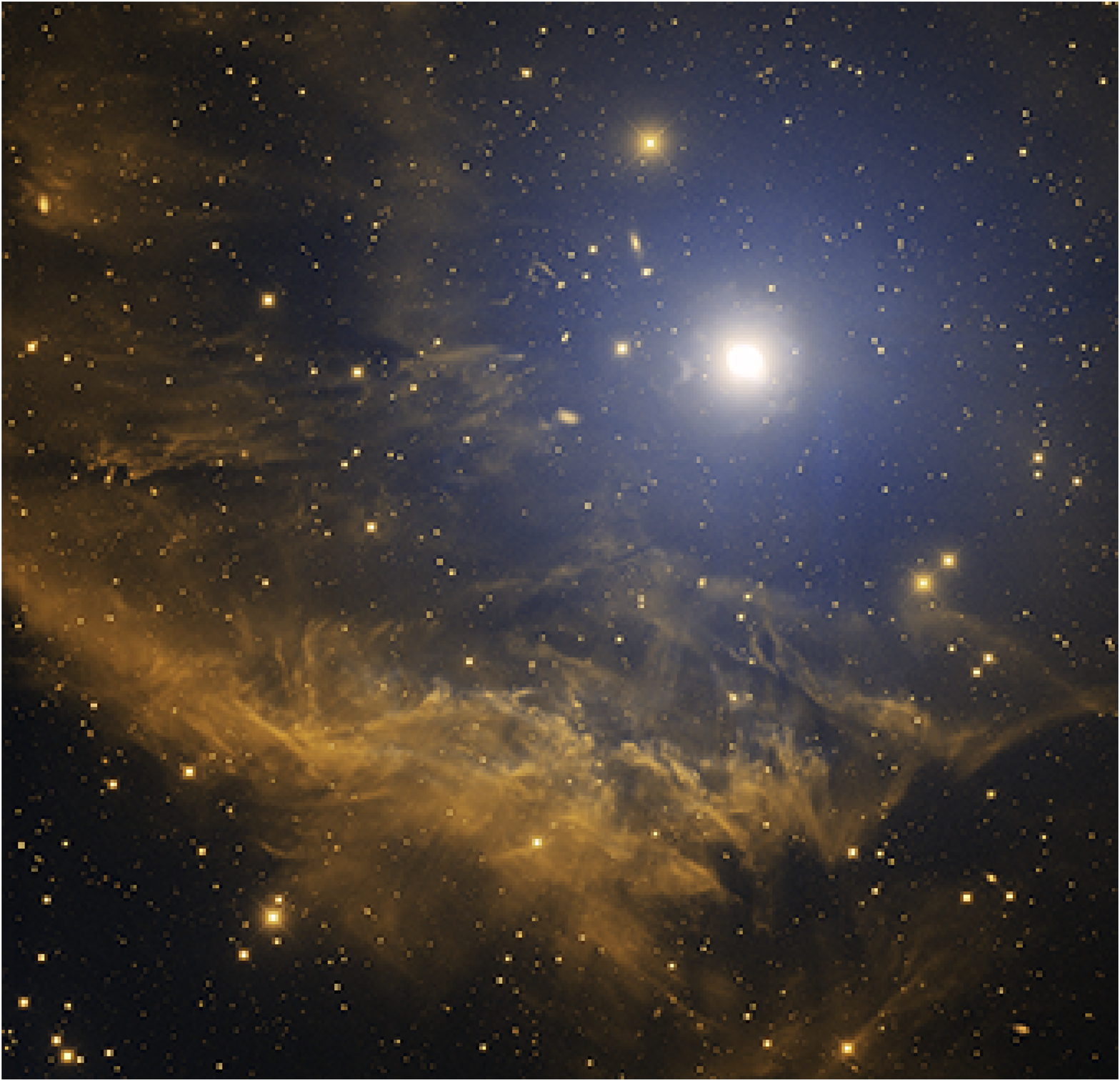}{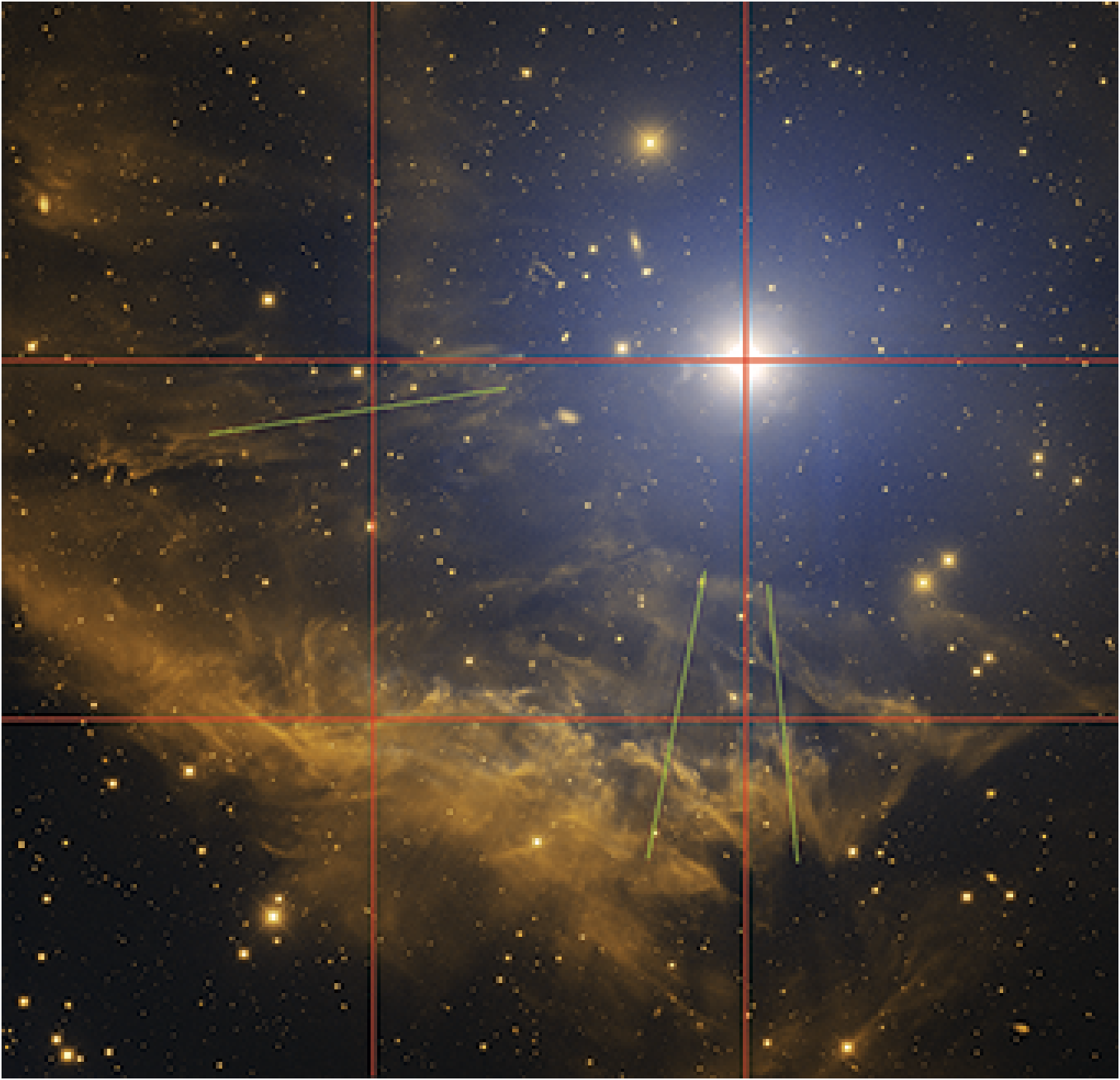} 
\caption{This image (left) of the planetary nebula NGC 3242 \citep{rector2006} was rotated and cropped in accordance to the rule of thirds (as shown by the red grid overlaid onto the right image). Structure in the nebula naturally produces several ``lines." Wisps in the nebula (marked in green in the right image) point back toward the bright core, making that a convergence point. This creates an illusion of depth, with the orange nebulosity feeling closer than the core. The illusion is enhanced by the warm/cool contrast, where cool colors, such as the blue halo around the core, feel more distant. Finally, the image is cropped so that the entire nebula is not seen. Structure in the nebula runs off of all the edges in the image, making the nebula feel large \citep{rector2015}.\label{fig:f5}}
\end{figure}

HST was also accidentally a pioneer in the use of orientations different than the traditional ``north is up, east is to the left" configuration of ground-based telescopes with equatorial mounts.  By virtue of the small FOV of the WFPC2 camera, HST was also unintentionally a pioneer in the use of tight croppings, a powerful tool that photographers have used for years. A tight cropping can create the illusion that an object is enormous, so big that the eye cannot capture it all. This is especially true in astronomical imaging, where the intrinsic size of an object is unclear because the object is unfamiliar.  A tight cropping that excludes a portion of the object in the image (i.e., the object or focal point extends beyond the borders of the picture) gives the sense that there is more to see, i.e., the image is showing only a tantalizing slice of a larger celestial landscape; e.g., Figure~\ref{fig:f5}.  Finally, a tight cropping can omit other objects in the field that would be distracting, focusing the viewer's attention on the region of interest.  That is not to say a tight cropping is always desirable. A wide field can help give perspective about an object or put it in context. Artists and photographers know this as the concept of ``negative space," where largely empty space surrounds the object of interest. When used properly, it can emphasize an object and draw the viewer's attention to it.  As with cosmetic retouching and other image manipulations, cropping and rotation should not be used to misrepresent; e.g., selectively cropping a field to remove objects that run counter to the scientific narrative.  For this reason it is recommended that raw, uncropped grayscale images for each dataset, as well as image processing information (e.g., filters and color assignments) be made available as ancillary information for all released images.

\subsection{Structure and Detail}

\begin{figure}
\figurenum{6}
\plottwo{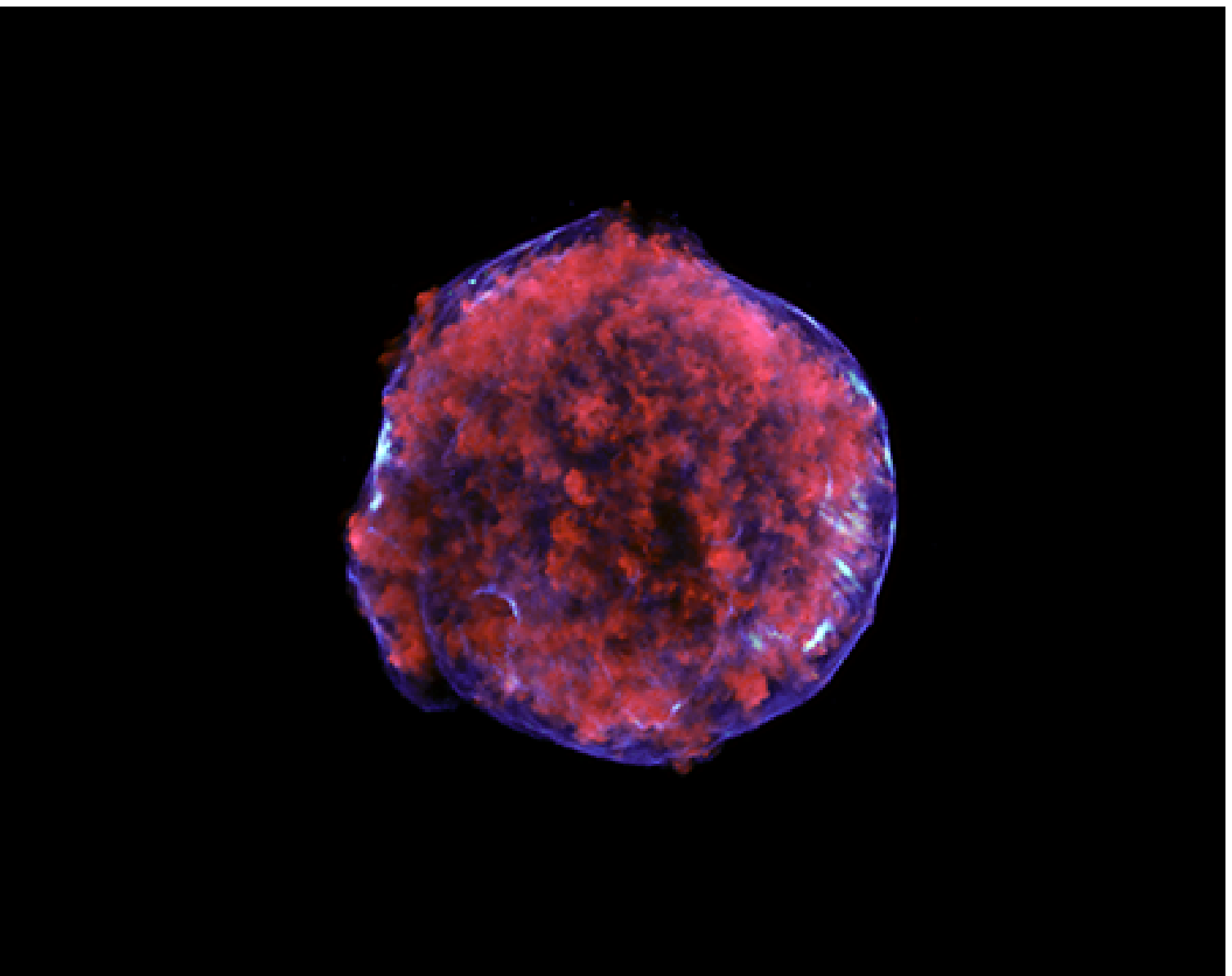}{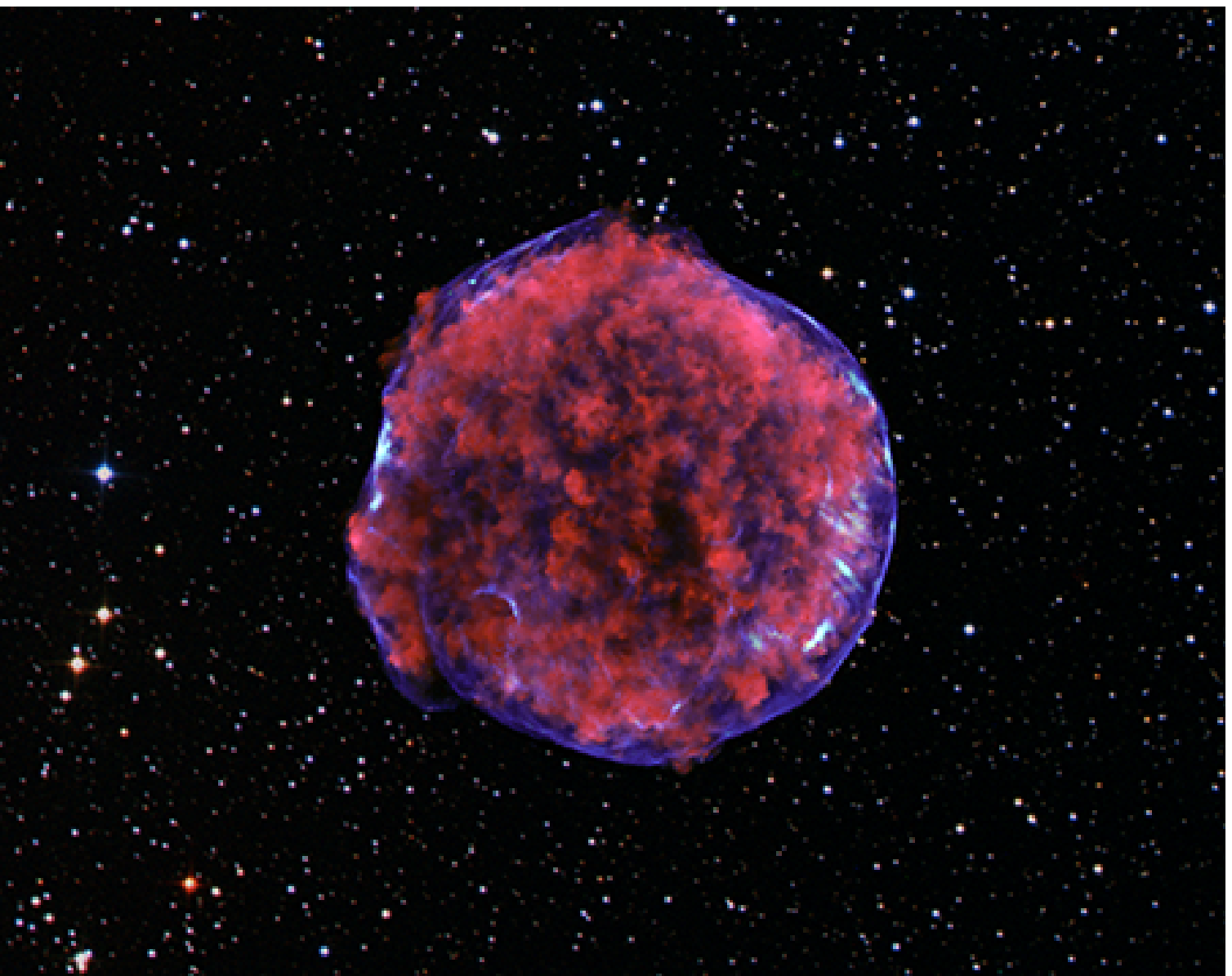}
\caption{A Chandra/ACIS image of the Tycho Supernova Remnant \citep{nasa2011b} as seen in X-rays only (left) and  optical DSS data added (right).  Because stars are not visible in the X-ray data, it is not clear that the image on the left is astronomical. The image on the right is therefore more naturally interpreted by a non-expert viewer \citep{smith2011}. \label{fig:f6}}
\end{figure}

Needless to say, what's actually in the image is the most important factor. Beauty is in the eye of the beholder of course, but there are some common characteristics that most aesthetically pleasing images have. Oftentimes, attractive pictures contain a mixture of the familiar and unfamiliar.  Familiar elements give the viewer a starting point for interpreting an image. For example, an important visual cue is the presence of stars, indicating that the image is astronomical \citep{smith2011}.  Figure~\ref{fig:f6} shows an example.  This can be especially useful if the image is in the news (or on a website) where astronomical images aren't necessarily expected. Images made with only radio or X-ray light often lack this visual cue, making interpretation more difficult.  It may seem contradictory, but having the presence of both familiar and unusual elements can help an image. The familiar helps to ground the viewing experience and give it context. The unfamiliar, on the other hand, gives you a reason to keep looking. It is as if your mind is saying, ``I've seen this before, but never quite like this!" \citep{rector2015}.

The visual aesthetic also depends on the type of astronomical object.  That's not to say every object of one class is more beautiful than the objects of another. But some types of objects have inherent advantages, at least when observed in the optical, because they naturally contain many of the visual elements we've discussed.

For example, some of the most beautiful objects in space are star-forming regions. Since nebular gas near embedded stars often contains sharp temperature gradients, narrowband filters can produce a wide range of colors.  Broadband filters can also differentiate between emission and reflection nebulosity. Embedded stars can also transform the gas, blowing away less dense areas and sculpting more dense regions. As a result, the structure inside the nebula can be diverse and complex. The combination of intricate structures and the range of colors can make star-forming regions truly stunning.  There's one additional important factor that aids the aesthetic appeal of star-forming regions in that they are externally lit by stars. This produces visual effects, such as shadows, that our mind uses to perceive structure and depth in a way similar to the way we do on a sunlit Earth. 

Planetary nebulae are also often beautiful objects because, like star-forming regions, they tend to have strong variations in temperature, creating a range of colors when observed through narrowband filters. The progenitor white dwarf also serves as an external light source. Planetary nebulae can have simple to highly complex structure. They also can be highly symmetric, which can improve their appeal; e.g., \citet{grammer1994}.  Similar to planetary nebulae, supernova remnants can be remarkably symmetric, especially when seen at X-ray energies. The also have complex and intricate details that can show dramatic variations in line emission.

Galaxies show a wide range in color due to the presence of a range of star temperatures as well as dust and HII regions.  Since galaxies are internally lit, it can be more difficult to perceive depth in galaxies. However, obscuration and reddening due to dust extinction give us visual cues that help us interpret the structure. Spiral galaxies usually have a wide range of colors and often have beautiful symmetry in their arms. However, breaks in this symmetry can also make a galaxy interesting, again a mix of the familiar and the unusual. A lack of young stars, dust, and complex structure usually makes elliptical galaxies less visually compelling. The structure of irregular galaxies is often so chaotic and unusual that they can lack visual cues that help the viewer to understand them. In contrast, globular clusters are highly virialized and show minimal variations in structure, making them easily identifiable to the casual observer, but somewhat less visually compelling.

Images that contain multiple types of objects can create compelling contrasts between the color and structure in the objects.  For example, images from the Hubble Frontier Fields galaxy clusters survey contain a wide range of galaxy types as well as curved arcs from gravitational lensing.  

\subsection{Anatomy of an Image: An Analysis of the Pillars of Creation \label{sec:pillars}}

\begin{figure}
\figurenum{7}
\plotone{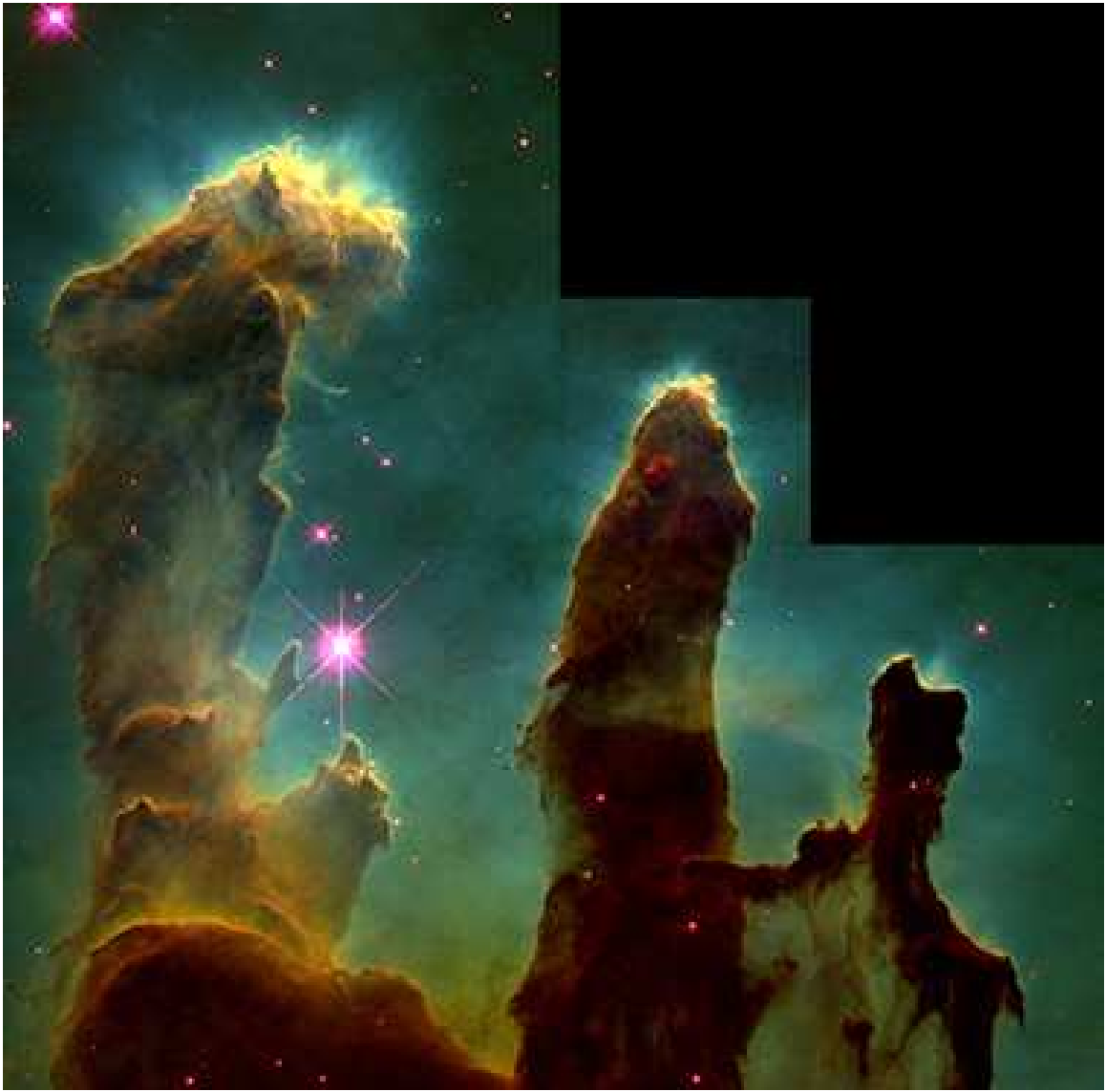}
\caption{The famous ``Pillars of Creation" image of M16 by \citet{hester1996}. Section~\ref{sec:pillars} describes several of the elements that make this image visually appealing.  \label{fig:f7}}
\end{figure}

With what we know about artistic principles of composition, we can look at the iconic HST ``Pillars of Creation" image of M16, shown in Figure~\ref{fig:f7}, and understand its popularity \citep{rector2015}.  First, the image is enhanced by the presence of several color contrasts. The image was made with \othree, \ha, and \stwo\ narrowband data assigned to blue, green, and red colors respectively (a chromatic ordering that amateur astronomers call the ``Hubble Palette.")  Subtle variations in temperature and density in the nebula cause significant differences in color, meaning the contrast of hue is abundant in the image. The light/dark contrast is visible between bright and dark regions in the pillars. The warm/cool contrast is seen between the cyan gas and the yellow pillars. This contrast gives depth to the image, causing the pillars to stand out against the backdrop. The complementary contrast is also present between pink (i.e., light red) halos of the stars and the cyan gas.

As a star-forming region, the pillars are also externally lit by several stars inside and out of the image, giving strong visual cues that help us to infer the shape of the pillars. For example, the large pillar on the left is illuminated by a star above the image, creating complex shadows underneath the top of the pillar. At the bottom of the pillar, the lighting makes it clear where additional ridges and ``knobs" are present. The middle and right pillars are illuminated from behind, causing the outer edges of the pillars to glow. The viewer can infer the right pillar as being in front because a portion of it overlaps the middle pillar.

The composition is compelling. The tight cropping caused by WFPC2's unique FOV gives the illusion that the pillars are large and seen close-up. The strong vertical lines in the pillars give them a sense of size and grandeur. However, curved lines within the pillars make them seem soft and pliable. The pillars also seem familiar, like those seen in canyons here on Earth. Whereas the green and yellow colors make them seem unfamiliar: a place like something on Earth, but not quite.

The image also benefitted from being one of the first of its kind. Using narrowband filters to produce images in this way was in 1995 a novel technique. As discussed before, it also greatly benefited from rapid distribution from a relatively new Internet. It was a perfect storm with a near-perfect image. Keith Noll, then the leader of the Hubble Heritage program, once said that its only imperfection is that the cosmetic defects in the image (e.g., the CCD bleeds and an internal reflection from the bright star near the center-left) were not removed \citep{noll2003}. Aside from that, it is an image that is worthy to be hung in an art gallery.

\section{Conclusion}

We have presented a review of the process of converting astronomical datasets into color images.  These images are important tools for scientific illustration and visualization as well as for public outreach. The quality of modern astronomical data, and technologies now available to manipulate them, allows for the creation of images in a way that exceeds the traditional definition of an astronomical image. With image-processing programs that use the layering metaphor, any number of astronomical datasets can be combined; and each dataset can be assigned any color. With this technique, images with as many as thirteen datasets have been produced. This is a fundamental shift from the traditional method of using only three datasets, and only with the primary colors of red, green, and blue. In the layering metaphor, each dataset can be individually scaled and colorized, creating an immense parameter space.

When creating these images we are in essence converting what a telescope can see into something we can see. It's a fundamental challenge because our telescopes observe objects that, with few exceptions, are invisible to our eyes. That is of course the reason why we build telescopes.   Using artistic principles of design and composition, we can create images that highlight the scientific structure within an image by using the color and intensity contrasts that the human eye uses to understand detail within an image. This strategy is particularly relevant for generating images with datasets outside of the optical window and for datasets with limited-wavelength coverage, e.g., emission-line optical filters. With visual grammar one can imply qualities that a two-dimensional image intrinsically does not show, such as depth, motion, and energy. However it is important to be aware of the differences between scientists and the public in how they interpret and understand images.  When done properly, these images can be scientifically illustrative as well as aesthetically pleasing for both audiences.

\acknowledgments

\end{document}